\documentstyle{article}

\def\be{\begin{equation}}
\def\ee{\end{equation}}
\def\bea{\begin{eqnarray}}
\def\eea{\end{eqnarray}}
\begin{document}

\pagestyle{empty}
\vskip-10pt
\hfill {\tt hep-th/0008161}
\begin{center}
\vskip 3truecm
{\Large\bf
On the holomorphically factorized partition function for abelian gauge theory in six dimensions
}\\
\vskip 2truecm
{\large\bf
Andreas Gustavsson
}\\
\vskip 1truecm
{\it Institute of Theoretical  Physics,
Chalmers University of Technology, \\
S-412 96 G\"{o}teborg, Sweden}\\
\vskip 5truemm
{\tt f93angu@fy.chalmers.se}
\end{center}
\vskip 2truecm
\noindent{\bf Abstract:}
We use holomorphic factorization to find the partition functions of an abelian two-form chiral gauge-field on a flat six-torus. We prove that exactly one of these partition functions is modular invariant. It turns out to be the one that previously has been found in a hamiltonian formulation.
\vfill
\vskip4pt
\noindent{August 2000, Revised September 2001}

\eject
\newpage
\pagestyle{plain}

\section{Introduction}
 
We study a two-form gauge-field $B_{MN}$ with self-dual field strength, on a six-dimensional manifold. In general there is not a unique partition function for the chiral two-form \cite{HNS}. Therefore it can not be described by any action in the usual sense (since that should have yielded a unique partition function). One way to compute its candidate partition functions is to holomorphically factorize the partition function for a non-chiral two-form gauge field. The (anti-)holomorphic coordinate one uses reflects the (anti-)chirality of the three-form field strength.

We start by briefly reviewing \cite{HNS}, where such a holomorphic factorization has been done. On any compact smooth orientable six-dimensional space we can introduce a symplectic basis of three-forms $E_A$ and $E_B$ with intersection numbers,
\bea
&&\int E_A\wedge E^t_A = 0, \int E_A\wedge E^t_B = -1\cr
&&\int E_B\wedge E^t_A = 1, \int E_B\wedge E^t_B = 0.\label{sympl}
\eea
Here $E_A$ and $E_B$ are infinite-dimensional column vectors whose elements are three-forms, and $E^t_A$ denotes the transponated vector. We let $[E^0_A]$ and $[E^0_B]$ denote the cohomology classes. This far everything has been defined without requiring a metric. When we introduce an Euclidean metric we get a complex structure in the space $\Omega$ of three-forms, defined via the Hodge operator *. As complex coordinates in $\Omega$ we use $Z = X + iY$, defined as
\be
E_B = X E_A + Y *E_A.
\ee
We also introduce the period matrix $Z^0 = X^0 + iY^0$ defined as
\be
E^0_B = X^0 E^0_A + Y^0 *E^0_A,
\ee
where $E^0_{A,B}$ are the harmonic representants in their corresponding (metric-independent) cohomology classes. The partition function $\cal{Z}$ for a chiral two-form $B$ is then given by
\be
{\cal{Z}}=\frac{\theta(Z^0|0)}{\Delta(Z)}\label{part fn}
\ee
provided that some non-holomorphic factors have cancelled, which we will discuss in a moment. Here $\theta$ is one of the Jacobi theta-functions which we have to pick. (To be able to pick the right one we must have more restrictions on the six-manifold \cite{W}). The quantum fluctuations contribute with the denominator $\Delta(Z)=\det{}^{'}(\Pi_A + Z \Pi_B)^{1/2}$. The prime $'$ indicates that the zero-modes are omitted from the determinant. The matrices $\Pi_A$ and $\Pi_B$ are defined as
\be
dF = \Pi^t_A E_A + \Pi^t_B E_B,
\ee
where $F$ is a basis for the space of two-forms modulo closed forms (i.e. an infinite dimensional vector whose elements are these two-forms). We note that the expansion-coefficient matrices $\Pi_A$ and $\Pi_B$ are constant over the manifold, and can be chosen to be independent of the metric (and hence of $Z$), since the $dF$ are orthogonal to the metric-dependent harmonics.

Now we will now consider the non-holomorphic factors in \cite{HNS}, and show that they cancel each other, i.e that
\be
\det{}^{'}{(Z-\bar{Z})}\det{(Z^0-\bar{Z}^0)} = 1.\label{non-holo}
\ee
(The prime $'$ indicates that we have removed the $Z^0$-part (i.e. the zero mode contribution) from $Z$). We then have to define (i.e. regularize) $\det{}^{'}{(Z-\bar{Z})}$. We use zeta-function regularization,
\be
\prod_{p=-\infty}^{\infty} a = \left.{a^{\sum_{p=-\infty}^{\infty}|p|^{-s}}}\right|_{s=0} \equiv a^{2\zeta(0) + 1} = 1,\label{zeta}
\ee
for $a$ being any constant number. The zeta-function fulfils $\zeta(0)=-\frac{1}{2}$. If we construct the symplectic basis by taking the harmonics and eventually some further three-forms to get $\left( ^6_3 \right)=20$ independent three-forms $E^{MNP}$ ($M,N,\ldots \in {1,\ldots 6}$), in the sense that $f_{MNP}(x)E^{MNP}=0$ (x is a point on the six-manifold) implies that the functions $f_{MNP}\equiv 0$, and then multiply these basis-forms by a complete set of functions (all functions, and the basis-forms are constrained by (\ref{sympl})) we have constructed a basis for the space $\Omega$ which gives
\be
Z_{p,q}(x)=Z^0_{ext}(x)\delta_{p,q}
\ee
where $Z^0_{ext}$ is a $20\times 20$-matrix which arises from the basis-forms, and which contains the contribution from the harmonics as one block. (On $T^6$ all the twenty three-forms $(E_A)^{mn}_{,0}$ and $(E_B)_{mn,0}$, corresponding to $p=0$, are associated with harmonics since the $3^{rd}$ Betti-number is $b^3(T^6)=20$ so in that case $Z^0_{ext}=Z^0$.) Zeta-function regularization now yields
\be
\det(Z-\bar{Z}) = \prod_{p=-\infty}^{\infty} \det(Z^0_{ext}-\bar{Z}^0_{ext}) = 1,
\ee
where we this time have not omitted the zero-modes (i.e. harmonic forms) from the determinant. Thus we have proved (\ref{non-holo}).

In the rest of this paper we will compute the determinant in (\ref{part fn}), which arose from the quantum fluctuations around the classical solution (to the classical equation of motion $d*H = 0$) in the path integral. We will restrict ourselves to a flat six-torus. We will find that this determinant is equal to the oscillator contribution in the hamilton formalism as obtained in \cite{DN}. We then prove that exactly one of the partition functions is modular invariant (i.e. invariant under the mapping class group $SL(6,Z)$ of the six-torus). We also show that if we happened to choose another symplectic basis, the partition functions just transform among themselves.

This is a revised version, where we also show (for a suitable choice of our symplectic basis) that $\theta\left[^{00\cdots 0}_{00\cdots 0}\right]$ is the only candidating theta-function as a modular invariant partition function on manifolds of the form $T_2\times M_4$, for $M_4$ being any compact, simply connected four-manifold. This gives a further motivation of the choice of theta-function made in \cite{B}.

\section{The computation}
We will now compute the contribution from the quantum fluctuations to the partition function for a chiral two-form, $B$, on a six-torus. We let $\theta^M = (\theta^m, \theta^6)\in [0,1]\times\cdots\times [0,1]$ ($m,n,... = 1,...,5$) denote the coordinates on the six-torus. As a symplectic basis of three forms we choose
\bea
(E_A)^{mn}{}_{,p} & = & e^{i2\pi p_R \theta^R}d\theta^m\wedge d\theta^n\wedge d\theta^6 \cr
(E_B)_{mn}{}_{,p} & = & e^{i2\pi p_R \theta^R}\frac{1}{6}\varepsilon_{mnMNP6}d\theta^M\wedge d\theta^N\wedge d\theta^P \label{sympl1}
\eea
and as a set for the space $F$ of two forms modulo closed forms we choose
\be
F^{mn}{}_{,p}   =   \frac{e^{i2\pi p_R \theta^R}}{2\pi i} d\theta^m\wedge d\theta^n.
\ee
We have then used the gauge fixing condition $F^{m6}=0$. Due to periodic boundary conditions we have $p \in Z^6$. We then find
\bea
(\Pi_A^t)^{rs}{}_{mn,p} & = & p_6 \delta_{mn}^{rs} \cr
(\Pi_B^t)^{rs}{}^{,mn}_{,p} & = & p_p \frac{G}{2} \varepsilon^{rspmn6}.
\eea
The matrix $Z$ for this choice of the symplectic basis is
\be
Z_{pq} = Z^0 \delta_{p,q}
\ee
where
\be
(Z^0)_{mn,rs} = \frac{1}{G^{66}}[\frac{1}{2}G^{6t}\varepsilon_{mnrst6} - i \frac{G_{m[r}G_{|n|s]}}{\sqrt{G}}].
\ee
This period matrix is computed in the appendix. We note that it belongs to (a subspace of) Siegel's upper half-plane ($Z=Z^t$ and Im $Z^{-1}>0$) and that it is conformally invariant. This should be so because the Hodge dual operator is conformally invariant and the form of our symplectic basis is independent of the metric.

We will now compute the determinant of
\bea
(\Pi_+)_{rs}{}^{mn}{}_{,p} & \equiv & (\Pi_A)_{rs}{}^{mn}_{,p} + (Z^0)_{rs,r's'}(\Pi_B)^{r's',mn}_{,p} \cr
& = & p_6 \delta_{rs}^{mn}+3 p_p \frac{G^{6t}}{G^{66}}\delta_{trs}^{pmn}-\frac{i}{2}p_p \frac{\sqrt{G}}{G^{66}}\varepsilon_{rs}^{}{}^{pmn6} \cr
& = & p_6 \delta_{rs}^{mn} - 3 p^p G_{6t} \delta_{prs}^{tmn} - \frac{i}{2} p^p \frac{g}{G^{66}\sqrt{G}}\varepsilon_{rsp}{}^{mn}
\eea
for each fixed $p \in Z^6$. In the last step we have used the relations (\ref{five metric}) and (\ref{epsilon}). We can linearly transform each $p\in Z^6$ separately
\be
p^p \rightarrow p \delta^p_1
\ee
in such a way that the distances are preserved,
\be
g_{pq} p^p p^q = g_{11} p^2,\label{length}
\ee
without changing the eigenvalues of $\Pi_+$. Then
\be
(\Pi_+)_{rs}{}^{mn}_{,p} = p_6\delta_{rs}^{mn} - 3 p G_{6t} \delta_{1rs}^{tmn} - \frac{i}{2} p \frac{g}{G^{66}\sqrt{G}}\varepsilon_{rs1}{}^{mn}.
\ee
We note that $(\Pi_+)_{1a}{}^{dc}_{,p}=0$, so in order to compute the determinant we only have to consider the two blocks $(\Pi_+)_{1a}{}^{1c}_{,p}=\frac{1}{2} \delta_{a}^{c} p_6$ and $(\Pi_+)_{ba}{}^{dc}{}_{,p}$ ($a,b,... = 2,...,5$). The determinant of the former block is an infinite product which however does not depend on the metric and so can be disregarded. In order to compute the determinant of the second block we determine its eigenvalues by making the ansatz
\be
V^{[ab]}_{cd} = A g_{p1} \varepsilon_{cd}{}^{pab} + B \delta_{cd}^{ab}
\ee
for the six eigenvectors, labeled by the multi-index [ab]. ($A$ and $B$ are some coefficients). Inserting this ansatz in the eigenvalue equation $\Pi_+ V=\lambda V$, identifying the coefficients in front of $\varepsilon_{cd1}{}^{ab}$ and $\delta_{cd}^{ab}$, and solving the secular equation, we find the eigenvalues
\be
\lambda_{\pm} = p_6 - p G_{61} \pm i p \frac{\sqrt{g g_{11}}}{G^{66}\sqrt{G}}
\ee
In order to see the $p_m$-dependence better, we transform back:
\be
\lambda_{\pm} = p_6 + p_m \frac{G^{m6}}{G^{66}} \pm i\sqrt{\frac{g^{mn} p_m p_n}{G^{66}}}\label{eigenvalue}
\ee
We have then used the formulas (\ref{length}), (\ref{five metric}) and (\ref{Cramer}) in appendix. The degeneracy of $\lambda_{\pm}$ is 3 for each sign.\footnote{There are 6 eigenvectors to $(\Pi_{+})_{ab}{}^{cd}$. But for each ansatz $V^{[ab]}$ we get {\it two} solutions, say $V^{[ab]}_{\pm}$, corresponding to the eigenvalues $\lambda_{\pm}$. These $6\times 2=12$ solutions must then be linearly dependent. It turns out that only 3 of the eigenvectors corresponding to $\lambda_{+}$ are linearly independent (and 3 corresponding to $\lambda_{-}$). To see this we note that if $V^{[ab]}_{+}$ and $V^{[cd]}_{+}$ are linearly independent, then so are also $V^{[ab]}_{-}$ and $V^{[cd]}_{-}$, which follows from that $g^{11}\neq 0$. We note that $g^{11}\neq 0$ because the metric tensor is a real and symmetric matrix and hence it can be diagonalized by an orthogonal matrix, say $S$. Thus $S^{-1}gS=S^tgS=$diag$((R_1)^2,\cdots,(R_5)^2)$. ($R_m$ is the radius of circle $m$ in the torus.) So $g^{11}=\sum_{m}S^{1}{}_{m}S^{1}{}_{m}/R_m^2>0$.} We then get
\be
\Delta(Z^0)=\prod_{p}{}'(\lambda_{+}\lambda_{-})^3=\prod_{p}{}'|\lambda_{+}|^6.
\ee
(The prime indicates that the product is over non-zero eigenvalues.) Written in this way it is manifestly real and positive.\footnote{It must be regularized in an $SL(6,Z)$-invariant way first though. This can be done as follows
\be
\Delta(Z^0)\equiv \left.\exp -\frac{\partial}{\partial s} E(s)\right|_{s=0}
\ee
where we define
\be
E(s)=\sum_{p}{}'|\lambda_{+}|^{-6s},
\ee
which converges when Re$(s)>1$. In the rest of the complex plane we define $E(s)$ by its analytic continuation. Then $E(s)$ has a simple pole at $s=1$ and is regular otherwise. The analytic function $F(s)\equiv E(s)(s-1)$ is real for real s for which Re$(s)>1$, which means that the (indeed analytic) function $\overline{F(\bar{s})}-F(s)$ is zero on this real half-line, and hence vanish identically by a fundamental theorem in complex analysis. Hence $F(s)$ is real on the whole real axis, and so $\Delta(Z^0)$ is real and positive.} It should be noted that we have only computed $\Delta(Z^0)$ in the subspace of Siegels upper half-plane. It is only in this subspace that we say that $\Delta(Z^0)$ is real and positive. By multiplying the eigenvalues in a different order one can recover the answer obtained in \cite{DN}, which however is not manifestly real and positive.

\subsection{Modular invariance}
Different choices of the symplectic basis should yield the same partition function, modulo the different candidate partition functions. The harmonic part of the symplectic basis is unique only up to a $Sp(10,Z)$-transformation. The large diffeomorphisms generate the mapping class group. On the six-torus this group is $SL(6,Z)$. It can be embedded in $Sp(10,Z)$ by noting that a diffeomorphism 
\be
d\theta^M\rightarrow U^M{}_N d\theta^N\label{SL6}
\ee
($U\in SL(6,Z)$) implies an $Sp(10,Z)$-transformation on the harmonic three-forms $E_{A,B}$ since $\int{d\theta^1\wedge\cdots \wedge d\theta^6}$, and thereby the symplectic product, is invariant under $SL(6,Z)$-transformations. Thus (\ref{SL6}) gives rise to a transformation in Sp(10,Z),\footnote{To avoid to much cluttering we will in the following consequently mean the harmonic forms $E_{A(B)}^0$ when we write $E_{A(B)}$.}
\be
\left( \begin{array}{c}
E'_A \\ E'_B
\end{array} \right)=\left( \begin{array}{cc}
A & B \\ C & D
\end{array} \right)
\left( \begin{array}{c}
E_A\\A_B
\end{array} \right)
\ee
and
\be
{Z^0}' = (C+DZ^0)(A+BZ^0)^{-1}.
\ee
Some symplectic transformations don't arise from any diffeomorphism. To study the effect of such transformations we keep  $\theta^M$ fixed (since the effect of changing the coordinates can be treated separately). These transformations can be chosen to be representatives of the equivalence classes in $Sp(10,Z)/SL(6,Z)$. By noting that $dF=\Pi_A^t E_A+\Pi_B^t E_B=(\Pi_A^t+\Pi_B^t \hat{Z})E_A=(\Pi_A^t+\Pi_B^t \hat{Z}^t)E_A=(\Pi_A+\hat{Z}\Pi_B)^t E_A=\Pi_{+}^t E_A$ (here $\hat{Z}\equiv X+Y*$) and, that under these representative transformations, $dF'=dF$ where ${E_A}'=A E_A+B E_B=(A+B\hat{Z})E_A$, we get
\be
{\Pi^t_{+}}'=\Pi^t_{+} (A+BZ)^{-1}.
\ee

The non-chiral partition function, ${\cal{Z}}(Z^0)\cdot\overline{{\cal{Z}}(Z^0)}$, is modular invariant (by which we mean invariant under transformations in the mapping class group, in this case $SL(6,Z)$) since it is defined from an action which is invariant under diffeomorphisms. Hence the chiral partition function transforms with at most a phase factor under modular transformations. Analyticity of the chiral partition function and reality of the phase implies that this phase must be a constant, i.e. independent of $Z^0$. (A real-valued analytic function must be a constant.)

We will temporarily consider manifolds of the form $M_6=T_2\times M_4$ where $M_4$ is simply-connected \cite{B}. We let $\{e^i\}$ be an ON basis of $H^2(M_4,{\bf{Z}})$ and $\{a,b\}$ a symplectic basis of $H^1(T_2,{\bf{Z}})$, $\int_{T_2}a\wedge b=1$. Further we let $Q^{ij}=\int_{M_4}e^i\wedge e^j$ be the intersection matrix of $M_4$. It fulfils $Q^{ij}=Q^{ji}$ and $Q^2=1$. We can then define a symplectic basis on $H^3(T_2\times M_4,{\bf{Z}})$ as
\bea
E_A^i & = & e^i\wedge b\cr \label{basis}
E_B^j & = & Q^{jk}e^k\wedge a.
\eea
The theta function $\theta\left[^{\phi}_{\theta}\right]$ can be determined from the holonomies \cite{W} $H(E_{A(B)})$ via the relations 
\bea
H(E_A^{i}) & = & (-1)^{2\phi^{i}}\cr
H(E_B^{j}) & = & (-1)^{2\theta^{j}}.
\eea
The mapping class group $SL(2,Z)$ of $T_2$ is generated by $T:a\rightarrow a+b$ and $S:a\rightarrow b$, $b\rightarrow -a$. These induce the following transformations of the symplectic basis of three-forms,
\bea
T:\left(
\begin{array}{c}
E_A \\
E_B
\end{array}
\right)
&\rightarrow&
\left(
\begin{array}{cc}
1 & 0\\
Q & 1
\end{array}
\right)
\left(
\begin{array}{c}
E_A \\
E_B
\end{array}
\right)\cr
S:\left(
\begin{array}{c}
E_A \\
E_B
\end{array}
\right)
&\rightarrow&
\left(
\begin{array}{cc}
0 & -Q\\
Q & 0
\end{array}
\right)
\left(
\begin{array}{c}
E_A \\
E_B
\end{array}
\right).
\eea
Invariance under $SL(2,Z)$ implies that the holonomies are constrained by
\be
H(E_B)=H(QE_A+E_B)\label{invT}
\ee
under the $T$-transformation, and by
\bea
H(E_A) & = & H(-QE_B)\cr\label{invS}
H(E_B) & = & H(QE_A)
\eea
under the $S$-transformation. Further these holonomies are always constrained by 
\be
H(E+F)=H(E)H(F)(-1)^{\int{E\wedge F}},\label{holonomies}
\ee
for any $E,F$ in the lattice $H^3(T_2\times M_4,Z)$, generated by $E_{A(B)}$ with integer coefficients. So we get first from the $T$-transformation
\be
H(Q^{ij}E_A^j+E_B^i)=H(Q^{ij}E_A^j)H(E_B^i)(-1)^{\int{Q^{ij}E_A^j\wedge E_B^i}}
\ee
which, by using (\ref{invT}), (\ref{invS}) and the fact that $Q^{ii}=0$, reduces to 
\be
H(E_B^i)=1,
\ee
and second from the $S$-transformation
\be
H(E_A^i)=H(-Q^{i1}E_B^1)H(-Q^{i2}E_B^2)\cdots H(-Q^{i,\frac{b_3}{2}}E_B^{\frac{b_3}{2}})=1.
\ee
This corresponds to the theta function $\theta[^{00\cdots 0}_{00\cdots 0}]$. 

We will now return to the flat six-torus. Then, under the $SL(2,Z)$ generated by $S$ and $T$, the theta-function found above transforms as follows (see e.g. \cite{LSW}),\footnote{The phase factor can be determined by expressing the spin-structure theta-function in terms of $D_1$ conjugacy class theta-functions. We find that $\theta\left[^{0 0 \cdots}_{0 0 \cdots}  \right]=\theta_{[00\cdots 0]}+(\theta_{[v0...0]}+\theta_{[0v...0]}+\cdots\theta_{[00...v]})+(\theta_{[vv0\cdots 0]}+$permutations$)+\cdots +\theta_{[vv\cdots v]}$. One knows how each conjugacy class theta-function transforms. The $SL(2,Z)$-generator $T=\left(\begin{array}{cc}
1 & 0\\
Q & 1
\end{array}\right)$ can be expressed in terms of the generators $\left(\begin{array}{cc}
1 & 0\\
E_k & 1
\end{array}\right)$ and $\left(\begin{array}{cc}
1 & 0\\
F_{kl} & 1
\end{array}\right)$ of $Sp(10,Z)$, where
\bea
(E_{k})_{ij}&=&\delta_{ij}\delta_{ik}\cr
(F_{kl})_{ij}&=&-\delta_{ij}(\delta_{ki}+\delta_{li})+\delta_{kj}\delta_{li}+\delta_{ki}\delta_{lj}
\eea
as follows:
\be
Q^{ij}=\sum_{k,l} (2Q^{kl}(E_k)^{ij}+Q^{kl}(F_{kl})^{ij}).
\ee
Under the $T$-transformation the $D_1$ conjugacy class theta-function $\theta_{[x_{1}\cdots x_{\frac{1}{2}b_3}]}$, with the conjugacy classes ($x_{i}$) being either (0) or (v), transforms with the phase factor
\be
e^{i \pi \sum_{k,l}Q^{kl}(-\|x_k-x_l\|+\|x_k\|+\|x_l\|)}=1.
\ee
where we have used $Q^{ij}=Q^{ji}$. This leads to (\ref{mod}) for the $T$-transformation. The $SL(2,Z)$ generator $S=\left(\begin{array}{cc}
0 & -Q\\
Q & 0
\end{array}\right)$ can be expressed in terms of the generators $\tilde{S}^i:(E_A)^i\rightarrow (E_B)^i$, $(E_B)^i\rightarrow -(E_A)^i$ and $T$ as
\be
S=(\tilde{S}T)^3\tilde{S}
\ee
where we have defined $\tilde{S}\equiv \tilde{S}^1\cdots \tilde{S}^{\frac{1}{2}b_3}$. This leads to (\ref{mod}) for the $S$-transformation.}
\bea
\theta\left[_{00\cdots 0}^{00\cdots 0} \right]({Z^0}')=\sqrt{\det(A+BZ^0)}\theta\left[_{00\cdots 0}^{00\cdots 0} \right](Z^0).\label{mod}
\eea
The flat six-torus corresponds to $M_4=T_4$. Now $T_4$ is not simply-connected, which implies that there are three-forms in $H^3(T_6,Z)$ which can not be written in the form (\ref{basis}). If we as a typical two-torus choose that with $a=d\theta^1$ and $b=d\theta^6$, then $d\theta^a \wedge d\theta^1 \wedge d\theta^6$ and $d\theta^a\wedge d\theta^b\wedge d\theta^c$ ($a,b,c,...\in {2,...,5}$) are those three-forms which are not in the form (\ref{basis}). However, under the modular group of this two-torus, these three-forms are invariant. On the remaining three-forms,
\bea
(E_A)^{ab} &=& d\theta^a\wedge d\theta^b\wedge b\cr\label{sympl2}
(E_B)_{ab} &=& \frac{1}{2}\varepsilon_{1abcd}d\theta^c\wedge d\theta^d\wedge a,
\eea
the $S$ and $T$ transformations act as
\bea
S:\left(
\begin{array}{cc}
(E_A)^{ab}\\
(E_B)_{cd}
\end{array}
\right)
&\rightarrow&
\left(
\begin{array}{cc}
0 & -\frac{g}{2}\varepsilon^{1abgh} \\
\frac{1}{2}\varepsilon_{1cdef} & 0
\end{array}
\right)
\left(
\begin{array}{c}
(E_A)^{ef}\\
(E_B)_{gh}
\end{array}
\right)\cr
T: \left(
\begin{array}{c}
(E_A)^{ab}\\
(E_B)_{cd}
\end{array}
\right)
&\rightarrow&
\left(
\begin{array}{cc}
\delta^{ab}_{ef} & 0\\
\frac{1}{2}\varepsilon_{1cdef} & \delta_{cd}^{gh}
\end{array}
\right)
\left(
\begin{array}{c}
(E_A)^{ef}\\
(E_B)_{gh}
\end{array}
\right).
\eea
For the transformations $S$ and $T$ we find that $\sqrt{\det(A+BZ^0)}$ is real and positive (being equal to, respectively, $(\lambda_{+}\lambda_{-})^{3/2}|_{p_M=(1,0,0,0,0,0)}$ and to $1$). The mapping class group $SL(6,Z)$ of $T_6$ can be generated by the $SL(2,Z)$ above together with transformations between various two-tori, such as $d\theta^1\rightarrow d\theta^2$, $d\theta^2\rightarrow -d\theta^1$, for which $\det(A+BZ^0)=1$ ($A$ becomes a permutatation and $B=0$). Since (\ref{mod}) holds also for such transformations (which is most easily seen by taking $Z^0$ to be a fixed point under permutations), we have that $\sqrt{\det(A+BZ^0)}$ is real and positive for any $SL(6,Z)$ transformation.

Since $\Delta(Z^0)$ also was found to be real and positive for any $Z^0$ related to some metric, we must thus have the transformation rule
\be
\Delta({Z^0}')=\sqrt{\det(A+BZ^0)}\Delta(Z^0)
\ee
under the mapping class group $SL(6,Z)$. Hence the partition function is modular invariant.

For the representative transformations in $Sp(10,Z)/SL(6,Z)$ we find
\bea
\Delta(Z^0)& = &\prod_{p\neq 0}\det(\Pi_{+,p})^{1/2} \rightarrow \prod_{p\neq 0}\det(\Pi_{+,p}(A+BZ^0)^{-1})^{1/2}\equiv \left( \prod_{p\neq 0}\det(\Pi_{+,p})^{1/2}\right) \prod_{p\neq 0}\det(A+BZ^0)^{-1/2}\cr
& = &\sqrt{\det(A+BZ^0)}\prod_{p\neq 0}\det(\Pi_{+,p})=\sqrt{\det(A+BZ^0)}\Delta(Z^0),
\eea
where we have applied (\ref{zeta}). Noting that the Jacobi $\theta$-functions transform in a similar way among each other, we have thus proved that the partition function is independent of which symplectic basis we choose, modulo the different candidate partition functions. Actually {\sl{each}} of the $2^{20}$ theta-functions corresponds to some choice of the symplectic basis $E_{A(B)}$.\footnote{By applying on the symplectic basis (\ref{sympl2}) the following $Sp(10,Z)$-transformation
\be
\left(
\begin{array}{c}
{(E_A')^{ab}}\\
{(E_B')_{cd}}
\end{array}
\right)
=
\left(
\begin{array}{cc}
\delta^{ab}_{ef} & (E_{ij})^{ab,gh} \\
0 & \delta_{cd}^{gh}
\end{array}
\right)
\left(
\begin{array}{c}
(E_A)^{ef}\\
(E_B)_{gh}
\end{array}
\right)
\ee
where $(E_{ij})^{ab,gh}=\delta^{ab}_{ij}\delta^{gh}_{ij}$, we get, from (\ref{holonomies}), that $H({(E_A')^{ij}}')=-1$ and all the other holonomies equal to $+1$. By then applying the $Sp(10,Z)$-transformation 
\be
\left(
\begin{array}{c}
{(E_A')^{ab}}\\
{(E_B')_{cd}}
\end{array}
\right)
=
\left(
\begin{array}{cc}
\delta^{ab}_{ef}-(E_{ij})^{ab}{}_{ef} & (E_{ij})^{ab,gh} \\
-(E_{ij})^{ab}{}_{ef} & \delta_{cd}^{gh}-(E_{ij})_{cd}{}^{gh}
\end{array}
\right)
\left(
\begin{array}{c}
(E_A)^{ef}\\
(E_B)_{gh}
\end{array}
\right)
\ee
we get instead $H({(E_B')^{ij}})=-1$ and all other equal to $+1$. By repeatedly applying these two kinds of $Sp(10,Z)$-transformations we can reach any theta function. Such transformations of course also changes the period matrix at the same time.}

\vskip 0.3truecm
I want to thank M{\aa}ns Henningson, Bengt E. W. Nilsson and Per Salomonson for discussions.

\newpage
\appendix
\section{Appendix - Computation of the period matrix }

There are in this paper some frequently used relations between the five-dimensional tensors (the metric $g_{mn}$ and $\varepsilon_{mnpqr}$, $m,n,...=1,...,5$) and the six-dimensional tensors ($G_{MN}$ and $\varepsilon_{MNPQRS}$, $M,N,...=1,...,6$. Here $\varepsilon_{123456}\equiv 1$):
\bea
G^{6r}g_{rt} & = & -G^{66}G_{6t} \label{five metric}\\
G\varepsilon^{rspmn6} & = &g\varepsilon^{rspmn}\label{epsilon}\\
G^{66}&=&g/G \label{Cramer}
\eea
where $G$ and $g$ are determinants of corresponding metrices.

The period matrix $Z^0=X^0+iY^0$ (which acts on three-forms in the isomorphic form $\hat{Z}^0=X^0+Y^0*$) on a six-torus can be obtained as follows:
\bea
&&d\theta^m \wedge d\theta^n \wedge d\theta^p \equiv \hat{\tilde{Z}}^{mnp}{}_{rs} d\theta^r \wedge d\theta^s \wedge d\theta^6\cr
&&\equiv\tilde{X}^{mnp}{}_{rs}d\theta^r \wedge d\theta^s \wedge d\theta^6 + \tilde{Y}^{mnp}{}_{rs} \frac{\sqrt{G}}{6}\varepsilon^{rs6}{}_{MNP}d\theta^M \wedge d\theta^N \wedge d\theta^P\cr
&&=(\tilde{X}^{mnp}{}_{rs}+\tilde{Y}^{mnp}{}_{uv} \frac{\sqrt{G}}{2}\varepsilon^{uv6}{}_{rs6})d\theta^r \wedge d\theta^s \wedge d\theta^6\cr
&& + \tilde{Y}^{mnp}{}_{uv} \frac{\sqrt{G}}{6}\varepsilon^{uv6}{}_{m'n'p'}d\theta^{m'} \wedge d\theta^{n'} \wedge d\theta^{p'}\cr
\Leftrightarrow\cr
&&\tilde{X}^{mnp}{}_{rs} + \tilde{Y}^{mnp}{}_{uv} \frac{\sqrt{G}}{2}\varepsilon^{uv6}{}_{rs6}  =  0\cr
&&\tilde{Y}^{mnp}{}_{uv} \frac{\sqrt{G}}{6}\varepsilon^{uv6}{}_{m'n'p'} = \delta^{mnp}_{m'n'p'}\cr
\Leftrightarrow\cr
&&\tilde{Y}^{mnp}{}_{uv} = -\frac{1}{2}\sqrt{G}\varepsilon^{mnp}{}_{uv}\cr 
&&\tilde{X}^{mnp}{}_{rs} = -3\delta^{mnp}_{rst}g^{tp'}G_{p'6}\cr
\Leftrightarrow\cr
&&(Z^0)_{mn,uv} \equiv \frac{1}{6}\varepsilon_{mnrst} \tilde{Z}^{rst}{}_{uv}\cr
&& = -\frac{1}{2}\varepsilon_{mnuvp}g^{pp'}G_{p'6}-i\frac{\sqrt{G}}{g}g_{m[u} g_{|n|v]}\cr
&& = \frac{1}{G^{66}}[\frac{1}{2}G^{6p}\varepsilon_{mnuvp6} - i \frac{G_{m[u}G_{|n|v]}}{\sqrt{G}}]
\eea


\newpage
\begin{thebibliography}{999}
\bibitem{HNS}M. Henningson, B. E. W. Nilsson, P. Salomonson, `Holomorphic factorization of correlation functions in (4k+2)-dimensional (2k)-form gauge theory', hep-th/9908107.
\bibitem{DN}L. Dolan, C. Nappi, `A modular  invariant partition function', hep-th/9806016.
\bibitem{W}E. Witten, `Five-brane effective action in M-theory', hep-th/9610234.
\bibitem{LSW}W. Lerche, A. N. Schellekens, N. Warner, `Lattices and Strings', Phys. Rep. 177 (1989) 1. Appendix C.
\bibitem{B}G. Bonelli, `On the supersymmetric index of the M-theory 5-brane and little string theory', hep-th/0107051.
\end{thebibliography}
\end{document}